# Differential privacy with adjustable preservation of multivariate relationships

Nicolas Ruiz

Universitat Rovira i Virgili
Departament d'Enginyeria Informàtica i Matemàtiques
Av. Països Catalans 26, 43007 Tarragona, Catalonia
nicolas.ruiz@urv.cat

**Abstract.** This paper derives an exact attenuation formula for multivariate cross-moments under differential privacy achieved through randomized response. From a proposed alternative formulation of randomization matrices compliant with differential privacy guarantees we show that, under attribute-level randomization and where the randomization of the whole distribution is characterized by Kronecker products, any cross-moment of the anonymized dataset involving distinct attributes is equal to the original cross-moment multiplied by attenuation parameters, themselves determined by the part of the overall privacy budget allocated to each attribute. This characterization enables explicit and transparent controls of multivariate relationships preservations in differential privacy through privacy budget allocation. A simple numerical illustration demonstrates that identical total budgets can yield substantially different covariance-preservation profiles. The results also make the multiplicative loss of higher-order interactions explicit and thus can support mechanism design guided by analytical priorities.

**Keywords.** Privacy, anonymization, differential privacy, randomized response

## 1 Introduction

Data on individual subjects are increasingly collected and exchanged. By their nature, they provide a rich amount of information that can inform statistical and policy analysis in a meaningful way. However, the richness of data is also a curse. Due to legal obligations and privacy concerns in societies, information is often not fully exploited to protect the confidentiality of respondents. In fact, such requirements shape the dissemination policy of micro data at national and international levels. The issue is how to ensure a sufficient level of data protection to meet releasers' concerns in terms of legal and ethical requirements, while offering users a reasonable richness of information. To manage this tension, anonymization stands out to reconcile both sides, by providing effective privacy protection while offering suitable data utility. A traditional approach to anonymization essentially consists of leveraging a repertoire of masking methods collectively known as statistical disclosure control (SDC, [1]). An SDC method with a heuristic parameter is run to anonymize the original data.

Another, and relatively more recent approach, is based on privacy models. A privacy model is an *ex-ante* privacy condition dependent on a parameter that guarantees a pre-specified upper-bound on disclosure risk, regardless of the impact on utility. Each privacy model can be enforced using one or several SDC methods whose parameters then depend on the privacy model parameter. There are currently two main families of privacy models, one based on *k*-anonymity [2] and the other on ε-differential privacy (DP) [3]. This paper will focus on the latter.

For all its current popularity, deploying DP in practice is not without difficulties. For example, the interpretation of the ε parameter or the complexity of implementation remain challenging [4]. But probably the most stringent issue is the resulting data utility that can be exploited from differentially private data. Indeed, if properly applied, DP can result in a severe degradation of data usefulness for analytical tasks [5].

**Contribution and plan of this paper**

There exists in the DP literature a very large set of contributions aiming at assessing, formalizing, estimating *etc.,* which quantities of interest, after application of DP, can be retrieved up to a reasonable degree [6]. In fact, the same applies to the literature on *k*-anonymity or even for SDC methods leaning or not against privacy conditions [1]. Note that we remain voluntarily vague on the terms "quantities of interest" and "utility guarantees", as further qualification is intimately linked to what a data user is in practice looking for in terms of data utility: means, variances, correlations, regression coefficients *etc.* To be precise, in much of the literature privacy is formulated as an *ex-ante* constraint on the data-release mechanism, while utility is typically specified separately, and often evaluated *ex-post*, through application-dependent measures such as information loss, distortion, or statistical accuracy [7]. Although utility-aware mechanism design can incorporate such objectives *ex-ante* (using for example optimal mechanism design [8]), the privacy definition itself generally does not prescribe which aspects of the data's statistical structure should be preserved.

In this paper, we focus on one, albeit quite central, aspect of data's statistical structure, that is multivariate relationships, of any order. Starting from a fairly general and popular application mechanism of DP guarantees, i.e. randomized response plus the use of Kronecker products for our multivariate setting, we first characterize in *Section 2* a new and simple alternative formulation of differentially private randomized response mechanisms. From this formulation, *Section 3* establishes the central result of this paper, by showing that any cross-order moments of a data set can be *ex-ante* preserved to a selectable extent. In practice, that means that a data curator can decide, before deploying DP, which relationships between attributes she wants to preserve, or lose, depending on the overall privacy budget that has been decided for the protection of the data set. To the best of our knowledge, this is the first time that such result is produced in the DP literature, and more generally in the privacy literature; after a thorough review we have not been able to identify contributions enabling a *precise* control of multivariate relationships *pre-anonymization*, for any approach. Finally, leveraging this new feature, *Section 5* presents a small empirical illustration, while conclusions and future research directions are gathered in *Section 6*.

## 2 Background elements

### 2.1 Randomized response

Let $X$ denotes an original categorical attribute with $1, \dots, r$ categories, and $Y$ its anonymized version. Given a value $X = u$, RR [9] computes a value $Y = v$ by using an $r \times r$ Markov transition matrix:

$$P = \begin{pmatrix} p_{11} & \cdots & p_{1r} \\ \vdots & \ddots & \vdots \\ p_{r1} & \cdots & p_{rr} \end{pmatrix}$$

where $p_{uv} = \Pr(Y = v|X = u)$ denotes the probability that the original response $u$ in $X$ is reported as $v$ in $Y$, for $u, v \in \{1, \dots, r\}$. To be a proper Markov transition matrix, it must hold that $\sum_{v=1}^{r} p_{uv} = 1 \; \forall u = 1, \dots, r$. $P$ is thus *right stochastic*, meaning that any original category must be spread along the anonymized categories. Moreover, if $P$ is nonsingular, it is proven in [9] that an unbiased estimator of the proportions of respondents whose true values fall in each of the $r$ categories of $X$ can be obtained.

The usual setting in RR is that each subject computes her randomized response $Y$ to be reported instead of her true response $X$. This is called the *ex-ante* or local anonymization mode. Nevertheless, it is also possible for a (trusted) data curator to gather the original responses from the subjects and randomize them in a centralized way. This *ex-post* mode corresponds to PRAM [10]. Apart from who performs the anonymization, RR and PRAM operate the same way and make use of the same matrix $P$.

RR is based on an implicit privacy guarantee called *plausible deniability* [11]. It equips the individuals with the ability to deny, with variable strength according to the parameterization of $P$, that they have reported a specific value. In fact, the more similar the probabilities in $P$, the higher the deniability. In the case where the probabilities within each column of $P$ are identical, it can be proved that perfect secrecy, or *perfect privacy*, in the Shannon sense is reached [12]: observing the anonymized attribute $Y$ gives no information at all on the real value $X$. Under such parameterization of $P$, that will denote hereafter $P^*$, a privacy breach cannot originate from the release of an anonymized data set, as the release does not bring any information that could be used for an attack. However, the price to pay in terms of data utility is high [11].

### 2.2 Differentially private randomized response

DP being a privacy condition but not a method in itself, it can be achieved through a variety of mechanisms. At a broad level, three foundational approaches are additive-noise mechanisms, including the Laplace and Gaussian mechanisms; randomized-response mechanisms, which directly perturb individual values; and probability-weighting mechanisms, such as the exponential mechanism, which privately select among discrete or non-numerical outputs. This taxonomy is not exhaustive, but it covers the principal building blocks used in many differentially private algorithms [6, 13].

Moreover, and as for RR, DP can be implemented under either a central or a local model. In central DP, individuals provide their original data to a trusted curator, who applies a privacy-preserving mechanism before releasing statistics or analytical outputs. In local DP, each individual randomizes their data before transmission, so the collector never observes the original values. Although the local model requires less trust in the collector, it generally introduces more noise and therefore yields lower statistical utility for a given privacy guarantee. As such many local DP-mechanisms became over the years designed based on RR [13]. It can also be noted that, during the inception of DP, randomized response was already considered as a method to produce differentially private data sets [3]. This is the choice that we will retain in this paper, remarking also that the results derived in what follows apply to both the central and local scenario.

For a categorical attribute with *1,…,r* categories, an DP-mechanism based on RR must have the following form:

***Proposition 1 (DP-RR, [14])****: The r×r matrix P of an ε-differentially private randomized response scheme is of the form:*

$$p_{uv} = \begin{cases} \frac{e^{\varepsilon}}{r-1+e^{\varepsilon}} \; if \; u = v \\ \frac{1}{r-1+e^{\varepsilon}} \; if \; u \neq v \end{cases} \quad with \; \varepsilon \geq \ln \max_{u=1,\dots,r} \frac{\max_{u=1,\dots,K} p_{uv}}{\max_{u=1,\dots,K} p_{uv}}$$

*Proposition 1* describes the constraints that DP happens to enforce on a randomization matrix, which incidentally must be symmetric and bistochastic.

Finally, and as the setting of this paper is inherently multi-dimensional, we need to endorse a multivariate approach to DP. Various ways are available in the literature, e.g. encoding-based mechanisms such as RAPPOR [15], pre-clustering of attributes [16], or using Kronecker products [17]. Here, we will adopt the latter. Multivariate randomized-response work going back decades observes that when independent RR is performed on each categorical variable of a data set, the resulting multivariate randomization matrix is the Kronecker product of the individual design matrices, and several recent contributions made use of it e.g. [18, 19]. Arguably, this is also an intuitive approach. Moreover, and coupled with the re-formulation of DP proposed below, it appears that this approach allows to control exactly the retention of dependencies in a protected data set.

## 3 An alternative formulation of randomization matrices for differential privacy

In this section we propose a simple, alternative formulation of randomization matrices for DP mechanisms. To do so, we consider a new class of randomization matrices recently introduced in the privacy literature, the $P(\lambda)$ class [20]. Considering again the case of a categorical attribute with *1,…,r* categories, this class is described by:

$$P(\lambda) = \lambda I + (1-\lambda)P^{*} \; with \; 0 \leq \lambda \leq 1 \qquad (1)$$

where $I$ denotes the identity matrix, and $P^{*} = \frac{1}{r}\mathbf{1}\mathbf{1}^{T}$ with $\mathbf{1}$ the all-ones vector.

Therefore, we have:

$$p_{uv}(\lambda) = \begin{cases} \lambda + \dfrac{1-\lambda}{r} \; if \; u = v \\ \dfrac{1-\lambda}{r} \; if \; u \neq v \end{cases} \qquad (2)$$

Under *Equation (1)*, a randomization matrix is built by considering two polar situations, no randomization at all (the identity matrix) or maximum randomization (the perfect privacy matrix $P^*$), while a single parameter $\lambda$ is weighing the two situations. The randomization matrix then follows. Arguably, it is a simple and intuitive way of deciding how randomization should be conducted, and also a straightforward procedure to generate randomization matrices.

Such randomization matrix reports the true category with a probability exceeding that assigned to each alternative category. Thus, the largest probability is given by the on-diagonal terms, while the smallest by the off-diagonal. Remark also that $P(\lambda)$ is a symmetric, bistochastic matrix, as *Proposition (1)*. Based on these elements, the following proposition holds:

***Proposition 2 (formulation of DP-RR as*** $\boldsymbol{P(\lambda)}$***)****: Let $M_\lambda$ be the randomization mechanism defined by Equations (1) and (2). Then $M_\lambda$ satisfies $\varepsilon$-differentially privacy for:*

$$\varepsilon(\lambda) = log\left(\frac{1+(r-1)\lambda}{1-\lambda}\right) \qquad (3)$$

***Proof****: For arbitrary categories $u, u', v \in \{1, \dots, r\}$, the likelihood ratio $\frac{Pr(M_\lambda(u)=v)}{Pr(M_\lambda(u')=v)}$ can take the values 1, $\frac{\lambda+\frac{1-\lambda}{r}}{\frac{1-\lambda}{r}}$ or $\frac{\frac{1-\lambda}{r}}{\lambda+\frac{1-\lambda}{r}}$. The maximum likelihood ratio is thus the second one:*

$$Max_{u,u',v}\ \frac{Pr(M_\lambda(u)=v)}{Pr(M_\lambda(u')=v)} = \frac{\lambda+\dfrac{1-\lambda}{r}}{\dfrac{1-\lambda}{r}} = \frac{1+(r-1)\lambda}{1-\lambda}$$

*Setting $\varepsilon$ equal to the logarithm of this quantity gives $\frac{Pr(M_\lambda(u)=v)}{Pr(M_\lambda(u')=v)} \leq e^{\varepsilon}$ for all $u, u', v \in \{1, \dots, r\}$, which proves $\varepsilon$-differentially privacy. The bound is attained by choosing v=u and any $u' \neq u$. It is therefore tight.*

Any randomization mechanism based on the $P(\lambda)$ class of randomization matrices appears to provide a direct characterization in terms of DP. Moreover, *Equation (3)* can be reversed, which gives:

$$\lambda(\varepsilon) = \frac{e^{\varepsilon}-1}{e^{\varepsilon}+r-1} \qquad (4)$$

Reinjecting this expression into *Equations (1)* and *(2)* characterizes an alternative formulation of randomization matrices for DP:

$$P(\varepsilon) = \left(\frac{e^{\varepsilon}-1}{e^{\varepsilon}+r-1}\right) I + \left(1 - \frac{e^{\varepsilon}-1}{e^{\varepsilon}+r-1}\right) P^*$$

with

$$p_{uv}(\varepsilon) = \begin{cases} \dfrac{e^{\varepsilon}}{e^{\varepsilon}+r-1} \; if \; u = v \\ \dfrac{1}{e^{\varepsilon}+r-1} \; if \; u \neq v \end{cases} \quad (5)$$

*Equation (5)* is in fact *Proposition 1*. Thus, DP can be formulated using randomization matrices that also consider explicitly the two polar situations mentioned above, no randomization at all *vs.* maximum randomization, with the weighting operating through the parameter $\frac{e^{\varepsilon}-1}{e^{\varepsilon}+r-1}$.

This result simply spells out an underlying matrix structure that happened to be implicitly present in *Proposition 1*, but which to the best of our knowledge has never been made explicit in the DP literature. However, one can note that it sheds an alternative light on the functioning of DP, specifically the interpretation of $\varepsilon$. *Figure* 1 displays *Equation (4)* for various sizes of categorical attributes, showing the implicit weight given to the identity matrix when selecting $\varepsilon$.

**Figure 1.** Implicit weighting of the identity matrix for various levels of $\varepsilon$ in DP.

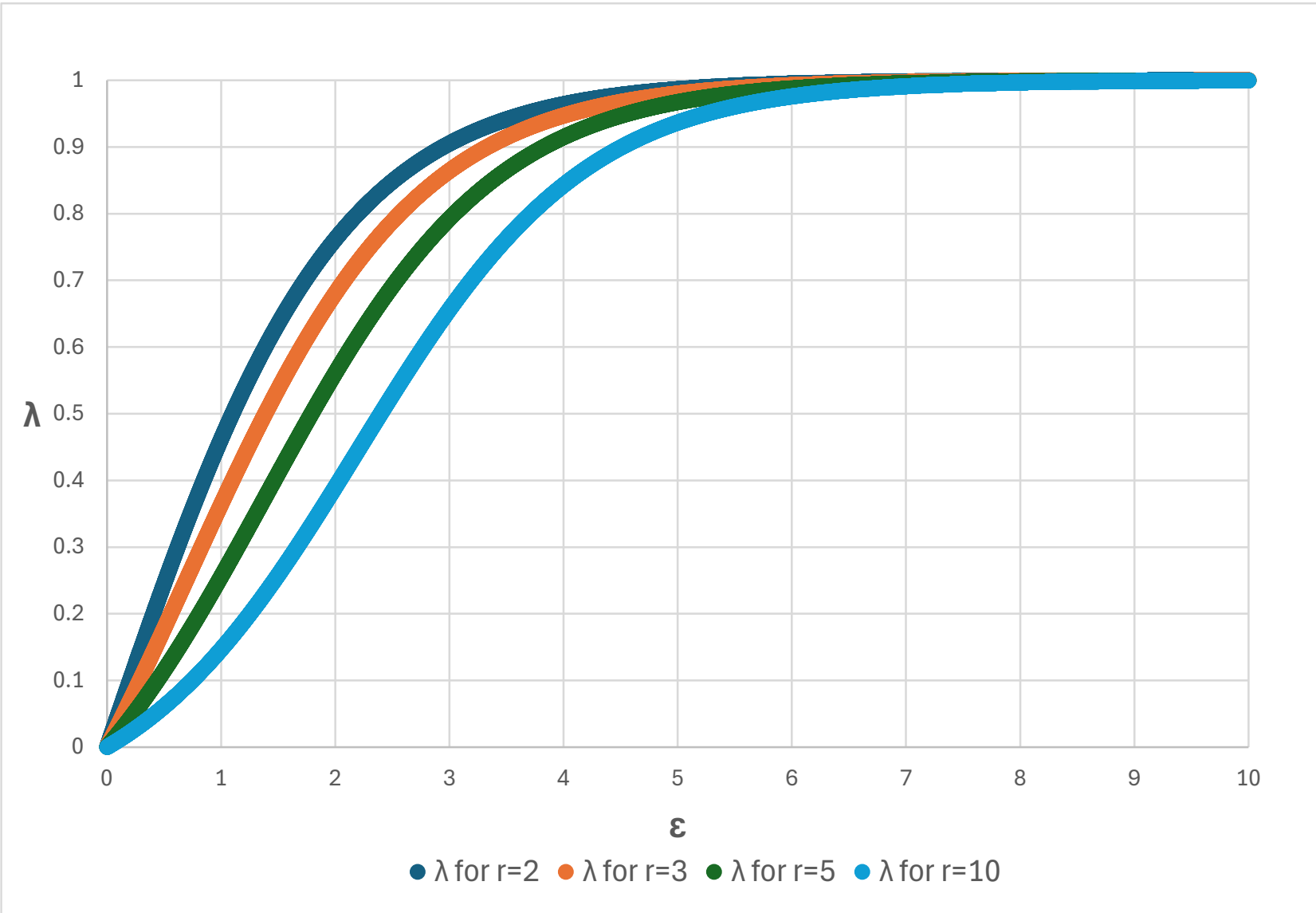

No formal rules exist for selecting $\varepsilon$. Some contributions in the literature recommend setting $\varepsilon$ to small values between 0.01 and 1 to enact protection, but it is known that practical deployments have required much larger values [5]. With the new formulation proposed, setting $\varepsilon$ below or in the vicinity of one is equivalent to performing no randomization at all at most 20% of the time, which is coherent with recommendations. However, when setting $\varepsilon$ above 3 it is equivalent to performing randomization with a weight attached to the identity matrix of 80% or more (except when the number of categories is large), which provides in practice very little protection, while above 4 it is as if no randomization is taking place.

## 4 Structural characterization of multivariate relationships attenuations in differential privacy

Consider a data set *D* with *1,…,M* attributes, each with $r_1,…,r_M$ categories, respectively, a record $x = (x_1, \dots, x_M)$, and let the corresponding attribute-level randomization matrix be as in *Equation (1)*:

$$P_m(\lambda_m) = \lambda_m I_{r_m} + (1 - \lambda_m)P^*_{r_m} \; with \; 0 \leq \lambda_m \leq 1 \; \forall m = 1, \dots, M \quad (6)$$

By *Proposition 2*, for attribute *m* the following holds:

$$\varepsilon_m(\lambda_m) = log\left(\frac{1 + (r_m - 1)\lambda_m}{1 - \lambda_m}\right) \quad (7)$$

Then, the randomization of the joint distribution of *D* can be represented by the Kronecker product of the matrices used for each attribute:

$$P_D(\lambda_1, \dots, \lambda_M) = P_1(\lambda_1) \otimes \dots \otimes P_M(\lambda_M) \quad (8)$$

The choice made in this paper to use the Kronecker product implies that the randomizations of each attribute are conditionally independent given *x*. Stated otherwise the randomization of one attribute does not depend on the randomizations of the others. To be clear, this assumption does *not* imply that the original attributes in *D* are assumed independent, nor that their randomized versions are independent. For example, $x_1$ and $x_2$ could be highly dependent and the Kronecker product will randomize them independently conditional on their original values, but the dependence is transmitted to $y_1$ and $y_2$, albeit in a way altered by DP. We mentioned above that there are many ways to enact DP in a multivariate scenario. But as will be clear with the theorem that follows, and beyond the computational simplicity of this choice, using Kronecker products for DP carries unintended benefits.

Now, for each attribute *m* define an arbitrary real-valued function $f_m$ of $x_m$ and of $y_m$ (the latter obtained from $x_m$ through application of DP), and define:

$$Z_m = f_m(x_m) \; and \; W_m = f_m(y_m)$$

and

$$\tilde{Z}_m = Z_m - E(Z_m) \; and \; \widetilde{W}_m = W_m - E(W_m)$$

From these elements, we have the following theorem:

***Theorem 1 (exact attenuation of multivariate relationships in DP)**: For every subset of distinct attributes* $S \subseteq \{1, \dots, M\}$, it holds that*:*

$$E\left(\prod_{m \in S} \widetilde{W}_m\right) = \left(\prod_{m \in S} \frac{e^{\varepsilon_m} - 1}{e^{\varepsilon_m} + r_m - 1}\right) E\left(\prod_{m \in S} \tilde{Z}_m\right) \quad (9)$$

***Proof**: For an attribute m, and conditional on the true value of the individual* $x_m$*, under the* $P(\lambda)$ *formulation DP retains* $x_m$ *with probability* $\lambda_m$ *and replaces it with a uniformly selected category with probability* $1 - \lambda_m$*, which implies:*

$$E[f_m(y_m)|x_m] = \lambda_m f_m(x_m) + (1 - \lambda_m)\frac{1}{r_m}\sum_{a=1}^{r_m} f_m(a)$$

*Taking the expectation of this last expression gives:*

$$E(W_m) = \lambda_m E(Z_m) + (1-\lambda_m)\frac{1}{r_m}\sum_{a=1}^{r_m} f_m(a)$$

*Next, we have:*

$$E(\widetilde{W}_m|x_m) = E(W_m|x_m) - E(W_m)$$

$$= \lambda_m f_m(x_m) + (1-\lambda_m)\frac{1}{r_m}\sum_{a=1}^{r_m} f_m(a) - \lambda_m E(Z_m) + (1-\lambda_m)\frac{1}{r_m}\sum_{a=1}^{r_m} f_m(a)$$

*Hence, we get:*

$$E(\widetilde{W}_m|x_m) = \lambda_m[f_m(x_m) - E(Z_m)] = \lambda_m \tilde{Z}_m \qquad (10)$$

*Now, consider any subset of distinct attributes S. By the assumed conditional assumption of the randomized variables $y_m$ given x, due to the use of the Kronecker product as the multidimensional setting retained, we have:*

$$E\left(\prod_{m\in S}\widetilde{W}_m\,|x\right) = \prod_{m\in S} E(\widetilde{W}_m\,|x)$$

*Using Equation (10), we get:*

$$E\left(\prod_{m\in S}\widetilde{W}_m\,|x\right) = \prod_{m\in S}\lambda_m\tilde{Z}_m = \left(\prod_{m\in S}\lambda_m\right)\prod_{m\in S}\tilde{Z}_m$$

*Finally, taking expectations with respect to x and using Equation (4), we have:*

$$E\left(\prod_{m\in S}\widetilde{W}_m\right) = E\left[E\left[\prod_{m\in S}\widetilde{W}_m\,|x\right]\right] = \left(\prod_{m\in S}\lambda_m\right)E\left(\prod_{m\in S}\tilde{Z}_m\right)$$

$$= \left(\prod_{m\in S}\frac{e^{\varepsilon_m}-1}{e^{\varepsilon_m}+r_m-1}\right)E\left(\prod_{m\in S}\tilde{Z}_m\right)$$

To illustrate the functioning of *Theorem 1*, consider the case of the covariance between two attributes $x_1$ and $x_2$ (with $\tilde{x}_1$ and $\tilde{x}_2$ their centered versions, respectively) both with the same number of categories *r*, and each randomized with $P_1(\lambda_1)$ and $P_2(\lambda_2)$, as defined in *Equation (6)*, to produce their randomized versions $y_1$ and $y_2$. We have:

$$\mathrm{cov}(y_1, y_2) = \frac{1}{r}(P_1(\lambda_1)\tilde{x}_1)^{\mathsf{T}}(P_2(\lambda_2)\tilde{x}_2) = \frac{1}{r}\tilde{x}_1^T P_1^T(\lambda_1)P_2(\lambda_2)\tilde{x}_2$$

$$= \frac{1}{r}\tilde{x}_1^T[\lambda_1\lambda_2 I_r + (1-\lambda_1\lambda_2)P_r^*]\tilde{x}_2 \qquad (11)$$

Then, because $\tilde{x}_1$ and $\tilde{x}_2$ are centered, it holds that:

$$\mathbf{1}^{\boldsymbol{T}}\tilde{x}_1 = \mathbf{1}^{\boldsymbol{T}}\tilde{x}_2 = 0$$

and thus:

$$P_r^*\tilde{x}_2 = \frac{1}{r}\mathbf{1}\mathbf{1}^T\tilde{x}_2 = 0$$

$$\tilde{x}_1^T P_r^* = 0$$

Simplifying *Equation (11)* and using *Equation (4)*, we have:

$$\mathrm{cov}(y_1, y_2) = \lambda_1\lambda_2\,\mathrm{cov}(x_1, x_2) = \left(\frac{e^{\varepsilon_1}-1}{e^{\varepsilon_1}+r_1-1}\right)\left(\frac{e^{\varepsilon_2}-1}{e^{\varepsilon_2}+r_2-1}\right)\mathrm{cov}(x_1, x_2)$$

Remark that, rather counter-intuitively, $Var(y_1) \neq \left(\frac{e^{\varepsilon_1}-1}{e^{\varepsilon_1}+r_1-1}\right)^2 Var(x_1)$. *Theorem 1*, formulated for a subset $S$ of distinct attributes, does not extend to the moments of any order of a single attribute. To see why, from the proof of Theorem 1 consider that:

$$E[W_m|x_m] = \lambda_m Z_m + (1-\lambda_m)\frac{1}{r_m}\sum_{a=1}^{r_m} f_m(a)$$

which implies, for the variance:

$$Var(E[W_m|x_m]) = \lambda_m^2 Var(Z_m)$$

But, by the law of total variance, we have:

$$Var(W_m) = Var(E[W_m|x_m]) + E(Var[W_m|x_m])$$

The second term on the right of the last expression represents the additional variability generated by randomization. It is a feature well-known in the RR literature [21], which cannot be omitted. This feature is however disappearing for cross-moments because the randomizations of two (or more) *distinct* attributes are conditionally independent, a consequence of the use of Kronecker products. Thus, *Theorem 1* characterizes the attenuation of *multivariate* relationships, while *univariate* moments behave differently because randomization adds its own variability.

More generally, *Theorem 1* establishes that, under the alternative parameterization of DP proposed in this paper, *any* cross-moment of a randomized distribution is linked to its original moment through an exact attenuation relationship, where the attenuation terms are the weights (themselves functions of the privacy parameters $\varepsilon$'s) allocated to the identity matrices. Probably the most direct consequence of *Theorem 1* is that it is possible to steer anonymization to precisely control for which multivariate relationship should be preserved to a lesser or greater extent. Indeed, through the choice of $\varepsilon$ for each attribute, one can control for the preservation of say certain covariances that are deemed more valuable (in terms of information value for data users) when close to the original ones in the randomized version of the dataset, while others less relevant could be more degraded.

But in doing so, one is confronted with the classic information/privacy trade off. Again, considering the covariance, large epsilons for two attributes will preserve it, while a small epsilon for at least one attribute will degrade the relationship, a rather intuitive consequence. Moreover, by choosing to largely preserve a covariance, this will directly weigh on an overall given privacy budget for the whole database, and thus other relationships will have automatically to be degraded. However, as unavoidable as the information/privacy tradeoff is, we believe *Theorem 1* allows to get a better appraisal of it.

In fact, those considerations are linked to the curse of dimensionality in privacy. This well-documented curse is the phenomenon that protecting a dataset becomes substantially harder as the number of attributes increases. Difficulties arise computationally, as the numerical operations required to protect a large dataset become costly if not intractable, but they also arise because statistical utility can deteriorate very quickly [16]. *Theorem 1* captures this curse on the statistical side, as the higher the order of a moment one wants to preserve, the more attenuation there is to deal with. For instance, and for a fourth-order cross-moment, 4 attenuation terms are involved. Thus, and as far as DP is concerned, high-order interactions disappear much faster, in fact multiplicatively so, than low-order effects.

Consequently, for a data set with many attributes multivariate relationships are mechanically almost lost, even for medium or high $\varepsilon$'s. Thus, the results in this paper establish an analytical hierarchy in the information/privacy trade off: for a given overall privacy budget, low-order relationships can be preserved more strongly, but high-order ones are attenuated more strongly, almost to the point of irretrievability. It must be noted that the ambition of this paper was not pretending to "solve" the curse of dimensionality. However, it provides a statistical manifestation of it, and potentially offers a solution to manage it by characterizing a particular structure in the preservation of multivariate relationships.

Finally, as mentioned above this paper considered the local and central scenario indifferently. However, for the latter those results can offer specific insights as to how an overall privacy budget should be allocated. Obviously, in the local scenario the data curator cannot know *ex-ante* what the strengths of the various relationships are in the original data. Thus, the curator must act blindly regarding how to allocate the privacy budget across the various attributes (a standard difficulty of the local scenario in privacy) or possibly act based on prior beliefs. However, in the central scenario the curator can witness the original strengths and thus decide how to dispatch the privacy budget. Indeed, using *Theorem 1* the curator can choose to strongly protect attributes which originally have poor relationships, as in that case attenuation will be harmless in terms of data utility, while relaxing protection on attributes with strong relationships, under the constraint of a given, overall privacy budget. Thus, for the central scenario the results developed in this paper offer room for improved management of the information/privacy trade off and the curse of dimensionality.

## 5 Numerical illustrations

In the example below, we act as a data curator who must protect a data set composed of five attributes $X_1, \dots, X_5$, each with four categories, so $r_m = 4$ (using the writings above). The curator sets a total privacy budget, under DP, of:

$$\varepsilon_{Tot} = \sum_{m=1}^{5} 5$$

Thus, for $r_m = 4$ we have the following attenuation coefficients:

$$\lambda_m = \frac{e^{\varepsilon_m} - 1}{e^{\varepsilon_m} + 3}$$

From the assigned total privacy budget, using basic sequential composition the curator designs three configurations:

- A balanced configuration where the budget is allocated equally: $\{1, 1, 1, 1, 1\}$
- A second where the preservation of the relationship between $X_1$ and $X_2$ is favored: $\{2, 2, 1/3, 1/3, 1/3\}$
- A third where priorities are graduated: $\{2, 1.5, 0.75, 0.5, 0.25\}$

Table 1 illustrates how the allocation of a fixed privacy budget determines the preservation of pairwise relationships. It reports percentages of the original covariances retained.

**Table 1.** Covariances preservations under different privacy budget allocation.

| Covariance between | Balanced | X_1 and X_2 favored | Graduated |
|---|---|---|---|
| X_1 and X_2 | 9% | 37.8% | 28.6% |
| X_1 and X_3 | 9% | 5.5% | 13.4% |
| X_1 and X_4 | 9% | 5.5% | 8.6% |
| X_1 and X_5 | 9% | 5.5% | 4.1% |
| X_2 and X_3 | 9% | 5.5% | 10.2% |
| X_2 and X_4 | 9% | 5.5% | 6.5% |
| X_2 and X_5 | 9% | 5.5% | 3.1% |
| X_3 and X_4 | 9% | 0.8% | 3.0% |
| X_3 and X_5 | 9% | 0.8% | 1.4% |
| X_4 and X_5 | 9% | 0.8% | 0.9% |

For the balanced configuration, all covariances retain the same proportion of their original values, as expected: the total budget is distributed evenly, so all pairwise relationships experience the same degradation. Still, with an $\varepsilon$ value of 1 for each attribute, i.e. when substantial protection is applied, only 9% of the original covariances is retained. Then, concentrating the budget on $X_1$ and $X_2$ substantially improves the preservation of their covariances (37.8%) compared to the balanced scenario, but almost cancels out relationships among the other attributes, a striking illustration of the trade-off involved. Finally, the graduated configuration produces an intermediate hierarchy of covariance preservation. Overall, an identical total privacy budget can induce very marked differences in data utility profiles, depending on how it is allocated across attributes.

## 6 Conclusions and future work

This paper considered the preservation of multivariate relationships under DP. Under a fairly general choice of application mechanism to enforce DP privacy guarantees, it first identified an alternative formulation for a differentially private randomized response scheme, as a combination of the identity matrix and the maximum, uniform randomization matrix. From this formulation, it then proceeded to an exact characterization of the attenuation of cross-moments involving distinct attributes, where attenuation is driven by the weights set to the identity matrix in the combination, and where those weights are determined by the attribute-level privacy budgets allocated from the overall budget.

Those results entail two main consequences. On the one hand, within the application mechanism for DP retained, the exact preservation of selected relationships can be specified before randomization, rather than assessed only after data release, a feature which to the best of our knowledge is new in the literature. The numerical illustration demonstrates that an identical total privacy budget can produce markedly different covariance-preservation profiles, depending on whether resources are distributed evenly or concentrated on selected attributes. Results of this paper therefore support the prioritization of analytical objectives by offering a better, more granular grasp of the information/privacy trade-off.

On the other hand, from those results it is clear that higher-order cross-moments accumulate more attenuation factors, making their preservation an arduous task. Thus, this paper exposes and analytically clarifies an unavoidable, structural source of utility loss in high-dimensional settings, by characterizing an exact mathematical relationship to appraise and manage what is otherwise known as the curse of dimensionality in privacy.

Several avenues for future research follow from these findings. A potential direction is to formulate the privacy budget allocation as an explicit optimization problem. Given a fixed total budget and a set of analytical priorities, one could seek allocations that maximize a weighted measure of cross-moment preservation or satisfy minimum retention requirements for selected relationships. Another direction concerns statistical estimation. The present results describe population-level attenuation, whereas practical applications rely on finite samples. Future work could then investigate estimators that correct for known attenuation, their sampling variability, and the conditions under which correction remains statistically useful. This would help distinguish the algebraic recoverability of a relationship from the precision with which it can be estimated, particularly when interaction orders are high.